\documentstyle[preprint,prl,aps]{revtex}
\begin{document}
\draft
\preprint{LAUR-95-950}
\title{The effect of inelastic processes on tunneling}
\author{Janez Bon\v ca$^{a,b,c}$ and S. A. Trugman$^a$}
\address{$^a$Theory Division and  $^b$Center for Nonlinear Studies,
Los Alamos National Laboratory,
Los Alamos, NM  87545, \\$^c$J. Stefan Institute,
University of Ljubljana, 61111, Slovenia}
\date{\today}
\maketitle
\begin{abstract}

We study an electron that interacts with phonons or
other linear or nonlinear excitations as it
resonantly tunnels.  The method we use is based on
mapping a many-body problem in a large variational
space exactly onto a one-body problem.
The method is conceptually simpler than previous Green's
function approaches, and allows the essentially
exact numerical solution of much more general problems.
We solve tunneling problems with
transverse channels,
multiple sites coupled to phonons,
and multiple phonon degrees of freedom and excitations.
\end{abstract}

We consider a single electron tunneling through a
resonant tunneling diode or a quantum dot, in the presence of
interactions with phonons or other excitations.
This interaction leads to
phonon assisted resonant tunneling \cite{tsui},
and affects the peak-to-valley current ratio, which is
important in device applications.
Most previous treatments of this problem use a Green's function
approach, often involving Keldysh formalism
\cite{scalapino,wingreen,glazman,cai,stoveng,zou,anda,lake}.
An exact solution was obtained only for the special case
of a single site coupled to a single phonon
mode in 1d \cite{wingreen,glazman,cai,stoveng}.
Experimentally important transverse degrees of freedom were
treated only to leading order in perturbation theory \cite{zou}.

The approach we use is to map the many-body problem in a
possibly large variational space exactly onto a one-body problem.  The
Schr\"odinger equation, for the most part in real space, is then
solved for the one-body problem.  In our view, this is a conceptually
simpler approach.  It also produces explicit solutions for broad
classes of problems that have not been solved before.  The solution
is essentially exact, in that the size of the variational space can
be systematically increased until the answer converges.
This approach need not make explicit use of Green's functions.
(If desired,
the Green's function can be recovered
from the wavefunction.)
Other inelastic tunneling problems can be solved by the same method.

One can solve essentially any problem where
a single electron tunnels, and where many-body interactions
are limited to a finite region of space.
The electron-phonon coupling may be nonzero on many
distinct sites, including several sites in the quantum well,
sites in the barrier, and sites in the leads near the barrier.
The electron may couple diagonally or off-diagonally
to many types of phonons, and multiple
quanta can be excited in the same or different modes.
The electron-phonon coupling may be nonlinear, and the
phonons may have nonlinear interactions among themselves.
The electron can have transverse degrees of freedom.
Arbitrary one-body interactions, including
barriers and disorder, can also be included.
The electron can interact in an arbitrary way,
including spin-flip scattering, with a group of interacting
``captive'' electrons in the tunneling region, so long
as the captive electrons cannot escape into the leads.
The method can also be used at non-zero temperature.
In practice, the method may make significant demands
on computer resources when more than about 10,000 inelastic channels
are included.

We consider the Hamiltonian
\begin{eqnarray}
H &=& H_{el} + H_{ph} + H_{el-ph} ,\label{ham}\\
H_{el} &=& \sum_j \epsilon _j c^\dagger _j c_j  -
\sum_{j,k} t_{j,k} ( c^\dagger _j c_k + h.c.),
\label{h_el}\\
H_{ph} &=& \sum_m \omega _m a^\dagger _m a_m ,
\label{h_ph}\\
H_{el-ph} &=& - \sum_{j,m} \lambda _{j,m} c^\dagger _j c_j
(a^\dagger _m + a_m).
\label{h_el-ph}
\end{eqnarray}
The potential $\epsilon _j$ on site $j$ can describe a tunnel barrier,
disorder, or a bias voltage.  The hopping amplitude $t_{j,k}$ can vary
from site to site.  $\lambda_{j,m}$ is the (diagonal)
coupling of an electron
on site $j$ to an optical or acoustic
phonon mode $m$.
A site can represent a single atomic
Wannier orbital or a larger region of space \cite{nonlinear}.

The method works for complicated barrier structures
and interactions described by the above Hamiltonian, and
for more general Hamiltonians.
To illustrate the method in a simple context, however,
we first
consider the case with
a single phonon mode that couples only to the electron density
on site 0.  The many-body problem is first restricted
to a variational subspace.  For illustrative purposes, only states containing
0, 1, or 2 phonon quanta are retained
in the example.  (A workstation could easily
handle thousands of states.)  The many-body scattering
problem in the variational
subspace is then mapped exactly
onto a 1-body problem with many channels, as shown in Fig. (\ref{prune}a)
and explained in the caption.
At zero temperature, an electron incident from the left is an incoming
plane wave on the lower left lead.  It has an amplitude
to exit on any of the six leads, corresponding to
elastic and inelastic backscattering and transmission.

We seek the solution of the Schr\"odinger equation
$E \psi _j = \sum _k H_{j,k} \psi_k$
on the tight-binding lattice of Fig. (\ref{prune}a)
with the known eigenvalue $E=-2t \cos (k_0) + \epsilon$,
where $t$ and $\epsilon$ are the hopping amplitude and diagonal
energy of the left lead, and $k_0$ is the incoming wavevector.
(The term ``lead'' refers to the translation invariant
part of the system.)
The boundary conditions are that an incoming wave is allowed
on only one lead.
The scattering problem is straightforward to solve.
One method is to ``prune''
any lead that has only an outgoing wave.
This exact procedure removes
the lead from the problem, while changing
the site energy $\epsilon _j$ on the last retained site
to a value that is in general {\it complex} and changes with $E$.
For example, the Schr\"odinger equation on site 0 is
$E \psi _0 = h_1 + \epsilon _ 0 \psi _0 - t_{0,1} \psi _1 $,
where $h_1$ includes the off-diagonal matrix elements
from site 0 to sites other than 1.
To prune the lower right lead in Fig. (\ref{prune}a),
note that there is a unique
outgoing wave of energy $E$ in the lead,
$\psi _ j = A \exp(i k_1 j)$ .
(If there are no propagating modes,
choose instead the decaying mode $\psi _ j = A \exp(- q_1 j)$.)
Back propagate this solution (through tunnel barriers
if necessary) to obtain
$\psi_0$ and $\psi_1$, using the Schr\"odinger equation
on each site $j$ to obtain $\psi _ {j-1}$
as a function of $\psi _j$ and $\psi _{j+1}$.
The new Schr\"odinger equation on site 0, with all the sites
in the lead removed, is now
$E \psi _0 = h_1 + \tilde \epsilon _ 0 \psi _0 $,
where $\tilde \epsilon _ 0 = \epsilon _ 0 - \alpha t_{0,1}$, and
$\alpha \equiv \psi_1 / \psi_0$ is generally complex.
The leads to be pruned in Fig. (\ref{prune}a) are marked
with vertical lines.  The system to be solved after
pruning is shown in Fig. (\ref{prune}b).

The problem in Fig. (\ref{prune}b) is so simple that it is best solved
by a recursive trick, which does not work in general.  To motivate the
general solution, consider the same problem, but where the electron at
site 0 interacts with two distinct phonon modes
of different frequencies.  Again for illustrative purposes,
choose a variational space that allows up to 2 phonon quanta in either
mode, for a total of $3 \times 3 = 9$ phonon states
(see Fig. \ref{prune}c).

The pruned problem is not solved as a standard eigensystem, since
the eigenvalue $E$ is known in advance.
Considering the amplitude $\psi_0$ to be known,
the problem is then to solve
a system of (complex) linear equations of the form
$A x = b$, where $x$ and $b$ are vectors, with $b$
proportional to $\psi_0$.  For this toy problem $A$ is an $8 \times 8$
matrix.
Once the system
$A x = b$ is solved, the Schr\"odinger equation on sites 0, -1, etc.,
is used to determine the wavefunction on the first two sites of the
left lead, and thus the coefficients $a_1$ and $a_2$ in $\psi _j = a_1
\exp(i k_0 j) + a_2 \exp(-i k_0 j)$.
The current $J$ leaving
through the pruned leads, corresponding to elastic or inelastic
transmission or backscattering in particular channels, is obtained
using
$J_{j \rightarrow k} = 2 ~ {\rm Im} ( \psi ^* _k ~ t_{k,j} ~ \psi _j)$.
The formula is applied for a retained site $j$ and a
neighboring pruned site $k$.
Current is conserved exactly,
globally and at each vertex.  This equation
can be used to calculate ordinary current
or a generalized current between
two many-body states \cite{green}.

Another form of electron-phonon coupling
modulates the hopping matrix element
$t$ rather than the on-site energy $\epsilon$,
\begin{equation}
H'_{el-ph} = - \sum_{j,k,m} \gamma_{j,k,m}
c^\dagger _j c_k (a^\dagger _m + a_m) + h.c.~~.
\label{h_el-ph2}
\end{equation}
This off-diagonal coupling represents the fact that when
an atom is displaced to the right, the hopping
amplitude $t$ to the atom on its right increases,
because it is closer.
A system with both types of electron-phonon coupling,
which can be solved by the same method, is shown in
Fig. (\ref{prune}d).

Figures (2-4) show essentially exact results for problems
that to our knowledge have not been previously solved.
Figure (\ref{trans1}) plots transmission for more than one site coupled
to a single phonon mode, and considers off-diagonal electron-phonon coupling.
Figure (\ref{manyph}) has coupling to many distinct phonon
modes of different frequencies \cite{manyref}.
Finally, Fig. (\ref{transverse}) considers transverse degrees
of freedom with electron recoil, which was previously
treated only to leading order in perturbation theory \cite{zou}.

As a simple test case, Fig.~(\ref{trans1}) shows
transmission through a quantum dot where a single phonon mode couples to an
electron on a site 0 with electron-phonon coupling strength
$\lambda$.
Hopping matrix elements are
$t_{k,l}=t_0$ between site 0 and sites $\pm1$
and $t_{k,l}=t$ for other nearest neighbors.
We model the weak coupling through a tunnel
barrier by a reduced $t_0$ in this paper, although we could
have just as easily used sites with increased $\epsilon_j$.
The same phonon mode
also modulates the hopping matrix element (see
Eq.~(\ref{h_el-ph2})).
A variational space with up to 8 phonon quanta
gives results accurate to the width of the plot lines.
The inset of Fig.~(\ref{trans1}) shows the
transmission for the case where phonons couple
only to the electron density, i.e. $\gamma=0$. Our result agrees
with previous calculations \cite{stoveng}.
The one-phonon sideband that we calculate was first seen experimentally by
Goldman et al. \cite{tsui}.
Note that the low-energy ``elastic'' peak and the
one-phonon sideband
(at $\omega \approx \omega_0-\lambda^2/\omega_0$)
are each composed of both elastic and inelastic transmission.
This is because there is an amplitude for
the electron in the left lead to couple to the first excited state
of the displaced harmonic oscillator for the electron on site 0,
and then to tunnel into the right lead
annihilating the phonon excitation,
leaving the (now undisplaced) harmonic oscillator in its ground state.
The calculation describes both inelastic tunneling, where
phonons are emitted, and polaron physics, where phonons
are emitted and reabsorbed.
Elastic and inelastic tunneling can be experimentally distinguished by
electroluminescence measurements \cite{teissier}.

Figures (\ref{trans1}a-d) show transmission when both diagonal
and off-diagonal electron-phonon coupling is present.
$\gamma_{ \{L,R \} }$ modulates the hopping between
site 0 and the \{left,right\} lead.
The only difference between figures (\ref{trans1}a-d)
is the relative sign of the coupling constants.
There are clearly dramatic interference effects,
which occur generically when a phonon is coupled to
more than one site.
The transmission peaks are much wider in
Fig.~(\ref{trans1}a) than in Fig.~(\ref{trans1}b).
This is attributed to the fact that the line width is
proportional to $\tilde t_0^{~2}$, where the effective coupling
$\tilde t_0 \approx t_0 + \gamma \langle x \rangle$.
The oscillator acquires a positive displacement $\langle x \rangle $
due to the $\lambda$ coupling.
Figures (\ref{trans1}c) and Fig. (\ref{trans1}d) represent
the same physical system reversed left to right.
The elastic part of the transmission is identical for
the two cases, while the total transmission is quite different.
This behavior is consistent with the unitarity requirements
for a time-reversal invariant system
with inelastic channels.
The total integrated transmission for (\ref{trans1}c) is quite different
from (\ref{trans1}d), because the sum rule \cite{wingreen}
is strongly violated for $\gamma$-type electron-phonon coupling.

Figure (\ref{manyph}) shows the transmission
through a single site quantum dot, which is coupled to 10 phonon modes
with different frequencies.  The electron-phonon coupling is
diagonal, as in Eq.~(\ref{h_el-ph}).
We compute transmission using variational
spaces that contain up to a total $N_{ph}=1,\dots,4$ phonon quanta
distributed in any way among the 10 modes.
Using subroutines for large sparse systems \cite{zlatev},
the system with $N_{ph}=4$ consisting of
$N_{st}=1000$ states (2000 channels, including
transmission and reflection) requires 20 seconds of CPU
time per energy point on a Sparc 10 workstation.
Each frequency point requires the solution of a
$N_{st}\times N_{st}$ sparse system of complex linear equations.
The inset of Fig.~(\ref{manyph}) displays
an enlarged portion of the transmission function in
the region of predominantly one-phonon contribution.
The weak electron-phonon coupling $\lambda=0.5$ is in
the experimentally relevant regime, where the single
phonon peaks dominate those due to multi-phonons.
Even so, it is clear that one does not get an
accurate description of inelastic tunneling in
a variational space containing only single phonon excitations,
and that $N_{ph}=3$ is required to achieve reasonable convergence.

It is straightforward to include transverse degrees of freedom, for
example to model the case where the tunnel barrier is an extended,
perhaps planar structure.
We investigated a model
containing $N_y$ parallel leads with periodic boundary conditions in
the (transverse) $y-$direction.
The Hamiltonian, written with real-space indices $j,~l$
in the $x-$direction and
momentum space indices $k,~q$ in the $y-$direction is
\begin{eqnarray}
H_{tr}&=&\sum_{j,k}(\epsilon_j-2t_{yj}\cos{k})
c^\dagger_{j,k}c^{ }_{j,k}
-\sum_{j,l,k}t_{j,l}(c^\dagger_{j,k}c^{ }_{l,k}+h.c.)\nonumber\\
&+&\sum_k \omega_k a^\dagger_{k} a^{ }_{k}
-\lambda/\sqrt{N_y}\sum_{k,q}c^\dagger_{0,k}c^{ }_{0,k+q}
(a^\dagger_{q}+a^{ }_{-q}),
\label{h_tr}
\end{eqnarray}
where the on-site energies are
$\epsilon_j=[\epsilon_l,\epsilon_0,\epsilon_r]$ for $[j<0,~j=0,~j>0]$
respectively.
Hopping matrix elements are
$t_{j,l}=t_x$ for nearest neighbor
$j,l\not= 0$, and $t_{j,l}=t_{x0}$ when $j=0$ or
$l=0$. Similarly $t_{yj}=t_y$ for $j\not= 0$ and $t_{yj}=t_{y0}$
otherwise. Diagonal electron-phonon coupling is restricted to sites where
$j=0$.
Due to the translational symmetry in the $y-$direction,
the total transverse momentum $k$ is conserved.  The electron momentum
changes only as
a consequence of the electron-phonon interaction.

Figure (\ref{transverse}) shows the transmission for
$N_y=6$ parallel leads.
Figure (\ref{transverse}) uses a variational space
with up to $N_{ph}=5$ phonon quanta in any phonon modes, which gives
satisfactory accuracy in the whole frequency and transverse momentum
range. For simplicity we use a dispersionless phonon
spectrum; however, generalization to
momentum-dependent phonon frequencies
and electron-phonon coupling is straightforward.
At small transverse momentum $k=0$ and
$\pi/3$, a strong nearly elastic resonance is located just below
the noninteracting resonance at $\omega(k)=\epsilon_0-2t_{y0}\cos{k}$.
Inelastic side-peaks at higher $\omega$
correspond to phonon creation,
usually accompanied by a change in the electron momentum.
For large transverse momentum $k=2\pi/3$ or $\pi$,
resonant states where the electron creates a phonon and
recoils to lower momentum
can have a lower energy than the state with no phonons.
This results in side-peaks at energies
below the weakened central peak.

{\it Generalizations:}
Finite temperature problems can be solved by having
the incoming electron arrive on different leads in
Fig. (\ref{prune}a) with the appropriate Boltzmann weights.
It should now be possible to model more realistic coupled
electron-phonon systems, to address such
questions as why the inelastic peak is observed experimentally
at the barrier phonon frequency rather than
that of the well \cite{tsui,mori}.
It would be interesting to study the stronger
electron-phonon coupling that arises when
electrons are localized on impurities, or from
phonon modes caused by crystal defects.
It should also be possible to model
more complicated band structures that include
several Wannier functions per unit cell,
and amorphous barriers if a suitable tight-binding
description is known.

{\it Acknowledgments:}
We thank
Jim Gubernatis, Selman Hershfield, Alan Katz, Roger Lake, Steve Lyon,
David Rabson, Darryl Smith, and John Wilkins
for valuable discussions.  This work was supported by the US DOE.

\begin{figure}
\caption[]{
(a) Each dot represents a basis state $|j,n \rangle$ in the many-body Hilbert
space.  The rows of dots are the sites $j=-3,\dots,3$ with
$n=0,1,$ or 2 phonon quanta.
The bonds represent non-zero off-diagonal
matrix elements in the Hamiltonian.  The horizontal bonds are
the hopping amplitudes $t_{j,k}$.
The vertical bonds represent
the electron-phonon interaction, with amplitudes
$ - \lambda $ on the lower and $- \lambda \sqrt{2}$ on the upper
vertical bond.
The dots can also be interpreted as Wannier
orbitals in an equivalent 1-body tight-binding model.
(b)  The system that results after pruning all but the incoming (lower left)
lead.  Sites with complex diagonal energies $\epsilon _j$ are shaded gray.
(c) The pruned version of the problem where an electron
at site 0 couples to two distinct phonon modes.  In the
$3 \times 3$ grid of gray dots, the vertical direction is
the number of quanta $n_1 = 0,1,2$ of type 1 phonons,
and the direction
into the page is the number of quanta $n_2$ of type 2 phonons.
Before pruning, there was one right and one left lead attached
to each gray dot.
(d)  A system with a single phonon mode, diagonal
electron-phonon coupling of the type in
Eq. (\ref{h_el-ph}) on site 0, and off-diagonal
electron-phonon coupling of the type in
Eq. (\ref{h_el-ph2}) between site 0 and sites $\pm 1$.
The pruned version is shown.}
\label{prune}
\end{figure}

\begin{figure}
\caption[]{ Transmission probability as a function
of the incident electron energy.  The heavy line
is the total transmission, and the lighter (lower) line
the elastic part.
A single phonon mode is coupled to the electron
density at site 0 with strength $\lambda$,
and to the electron hopping amplitude between site 0
and sites $\pm 1$ with strength $\gamma$.
The other parameters of the
Hamiltonian are:  $\epsilon_l=\epsilon_r=\epsilon_0=0.0$
(no voltage drop across the dot),
$t_0=0.2$, $\omega_0=1.0$ and $N_{ph}=8$.
All energies are in units of the
hopping $t$ in the leads. }
\label{trans1}
\end{figure}

\begin{figure}
\caption[]{ Transmission through a single site
that is coupled to ten phonon
modes with phonon frequencies $\omega_n$ uniformly distributed between
$\omega_{min}$ and $\omega_{max}$, with electron-phonon coupling
constants $\lambda_n=0.5\omega_n$. Different curves represent runs
with basis sets containing a different total number of phonon quanta
$N_{ph}$. }
\label{manyph}
\end{figure}

\begin{figure}
\caption[]{ Transmission through $N_y=6$ transverse channels
as a function of the
energy $\omega$ of the incident electron at four different choices of
transverse momentum $k$, with $t_{x0}=0.2$.
The upper and lower lines represent total and elastic
transmission respectively. }
\label{transverse}
\end{figure}


\begin{references}
\bibitem{nonlinear}
Nonlinear phonon modes or nonlinear interactions
among phonons can be modeled by adding terms of
higher than quadratic order to $H_{ph}$,
such as $(a^\dagger)^4 + h.c$.
Nonlinear electron-phonon coupling could
be included with terms such as  $c^\dagger  c (a^\dagger)^2 + h.c$.
The same method works with nonlinear terms included.
\bibitem{green} Once the many-body problem has been mapped onto
a one-body problem as shown in Fig. (\ref{prune}a),
the reader who prefers can use Green's function
methods at this point,
rather than solving for the wavefunction.
The change in energy $-\alpha t_{0,1}$ is then the self energy.
\end{references}
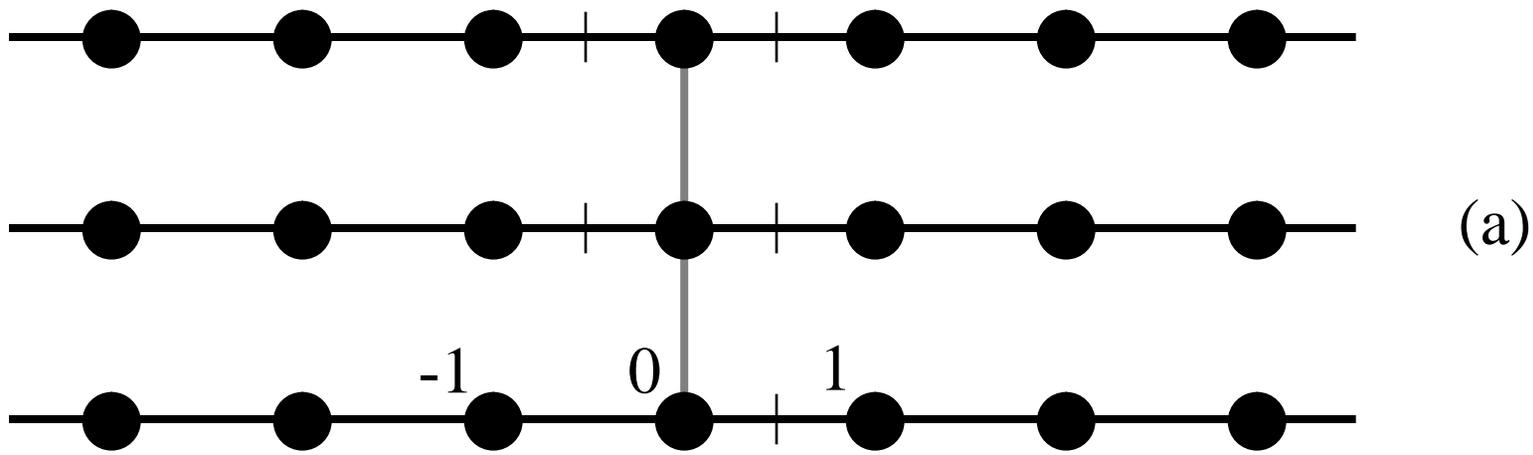
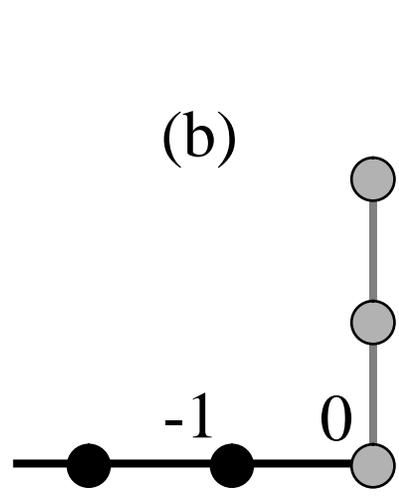
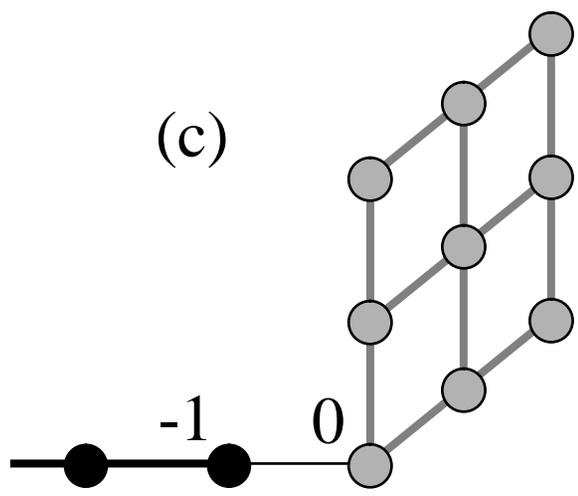
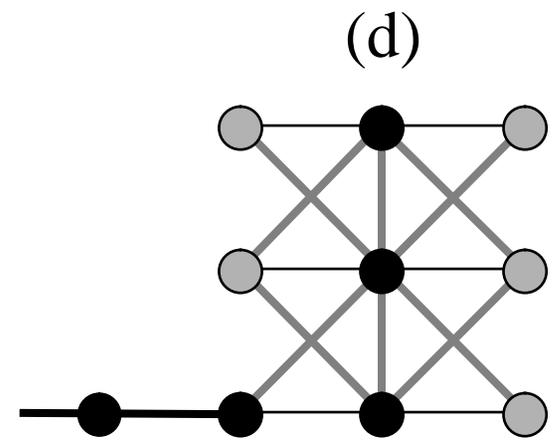
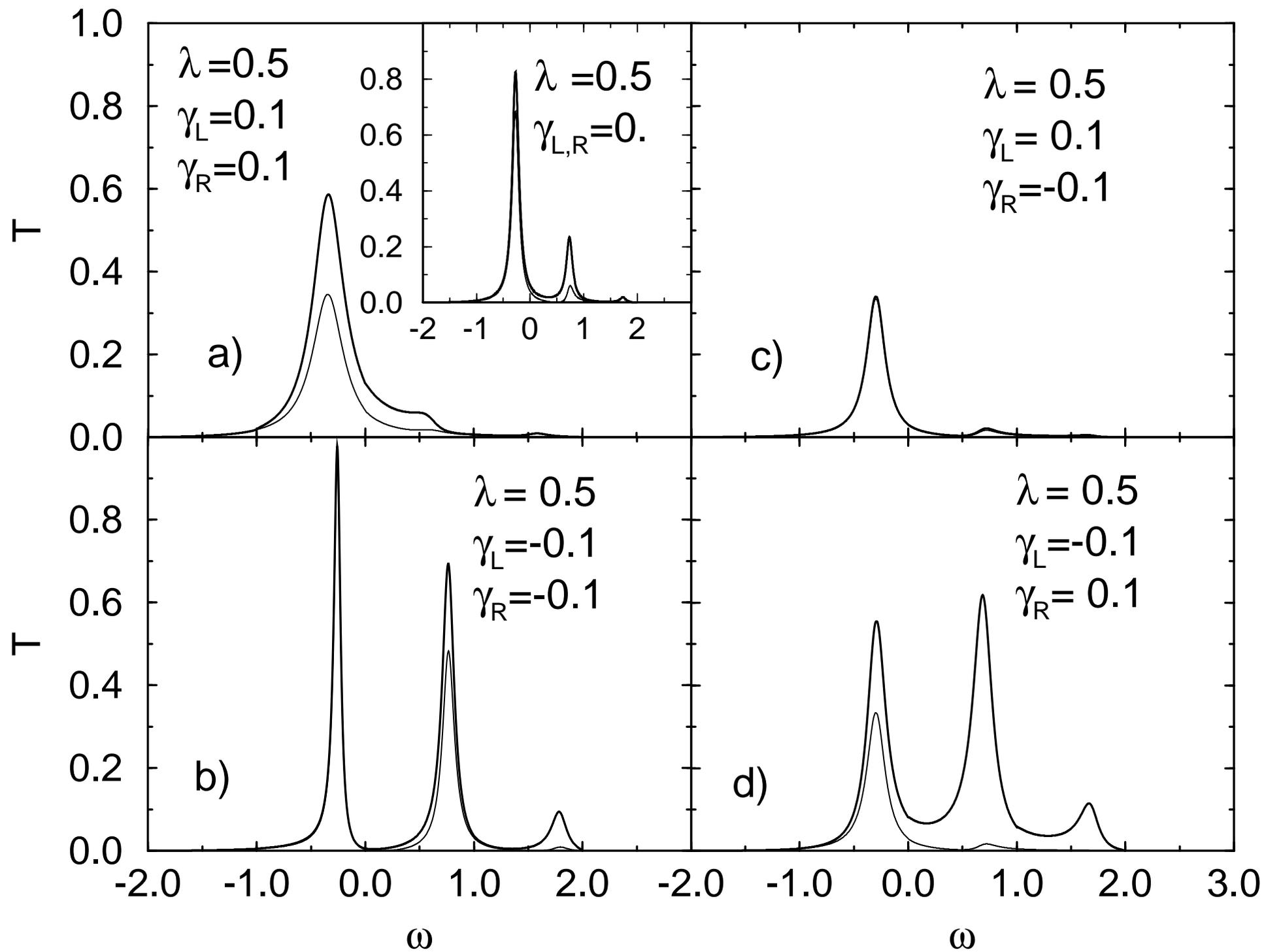
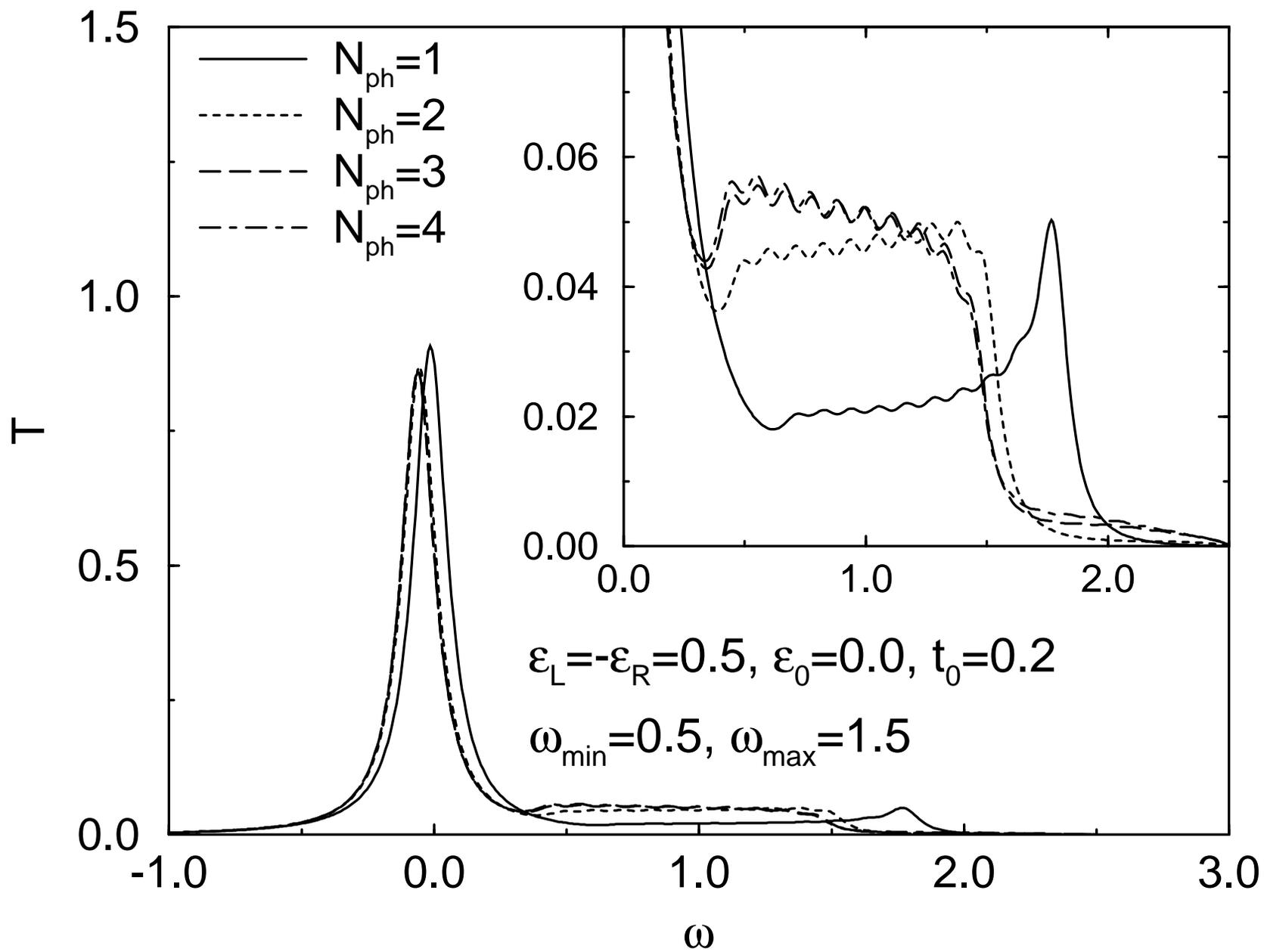
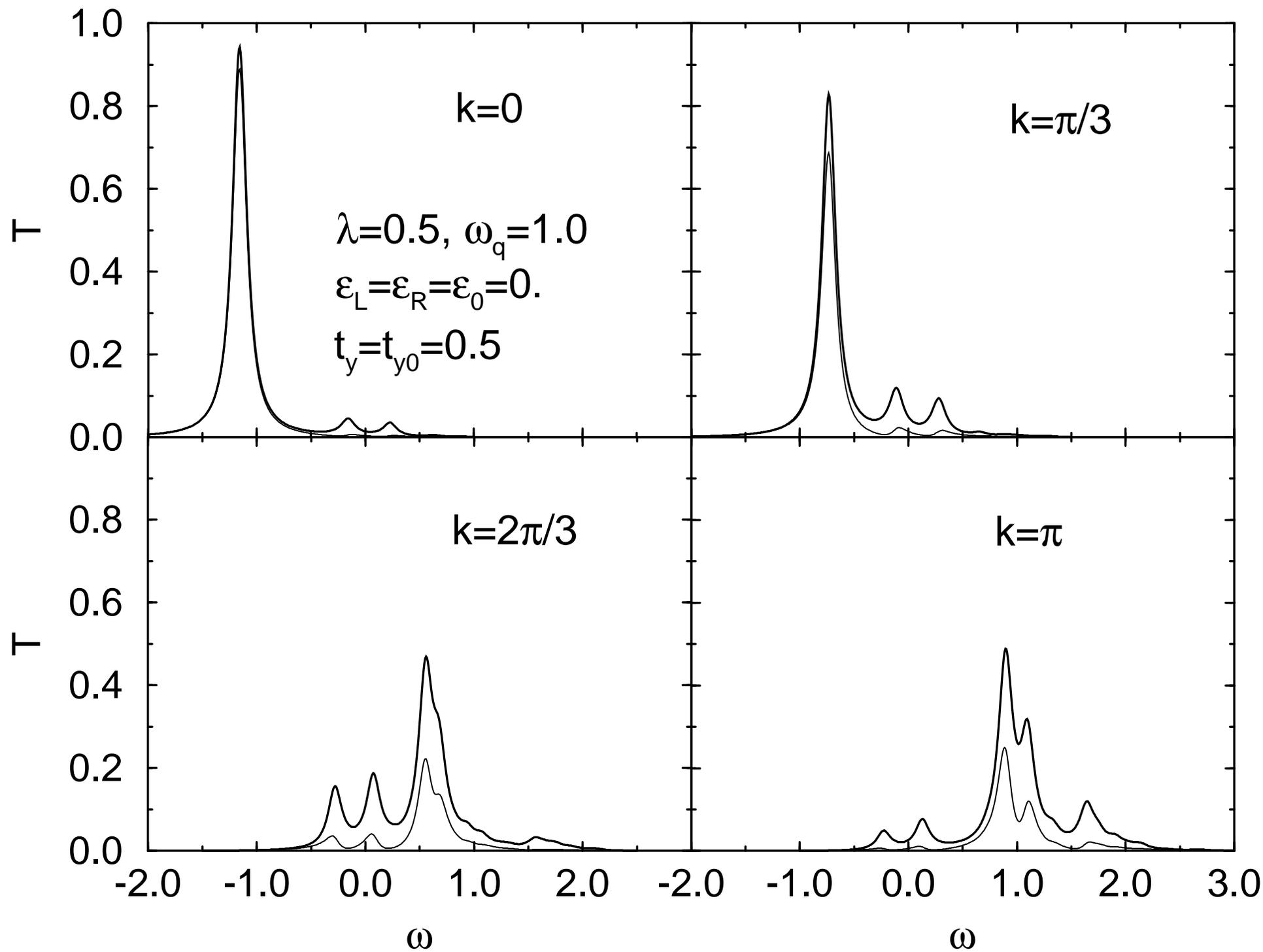
\end{document}